\documentclass[onecolumn,prb]{revtex4}

\usepackage{epsfig,amssymb,amsmath}

\begin{document}

\title{Common features of vortex structure in long exponentially shaped Josephson
junctions and Josephson junctions with inhomogeneities}

\author{T.L. Boyadjiev~$^{1}$}
\author{E.G. Semerdjieva~$^{2}$}
\author{Yu.M.Shukrinov~$^{1,3}$}

\address{
$^{1}$Joint Institute for Nuclear Research, 141980 Dubna, Russia\\
$^{2}$Plovdiv University, 24 Tzar Asen Str., Plovdiv, 4000, Bulgaria \\
$^{3}$Physical Technical Institute, Dushanbe, 734063, Tajikistan}

\begin{abstract}
We study vortex structure in three different models of long Josephson junctions:
exponentially shaped Josephson junction and Josephson junctions with resistor and shunt
inhomogeneities in barrier layer. Numerical calculations of the possible magnetic flux
distributions and corresponding bifurcation curves have done. For these three models the
critical curves ``critical current-magnetic field'' are constructed. We develop an idea
of the equivalence of exponentially shaped Josephson junction and rectangular junction
with distributed inhomogeneity and demonstrate that at some parameters of shunt and
resistor inhomogeneities at the ends of the junction the corresponding critical curves
are very close to the exponentially shaped one. Pacs: {05.45.+b}{74.50.+r} 74.40.+k.

\end{abstract}

\maketitle

An exponentially shaped Josephson junction (EJJ) has been recently suggested as a
tunable flux-flow oscillator operating at frequencies above 100 GHz \cite{benab96}. The
exponential variation of the junction width provides better impedance matching with an
output load  and allows one to avoid the chaotic regimes inherent to rectangular
junctions \cite{cmc_02}.

In this paper we study the stability of the vortices in long Josephson junctions (JJ)
and develop an idea \cite{sbs_05} that vortex structure in exponentially shaped JJ and
in JJ with inhomogeneity at the applicable end of the junction have common features. The
use of inhomogeneities might have some technological advantages for construction of flux
flow oscillator.

In order to investigate the stability of bound states we solve the static non-linear
boundary value problem together with corresponding Sturm-Liouville problem
\cite{bpp_jinr88}, \cite{tlb_02}. The minimal eigenvalue of the Sturm-Liouville problem
allows to make a conclusion about the stability of the corresponding vortices in JJ
\cite{sbs_05} -- \cite{bss_06}.

The corresponding boundary value problem for static dimensionless magnetic flux
$\varphi(x)$ in case of in-line geometry has the form
  \label{inline}
    \begin{gather}
        -\varphi_{xx} + j_C (x)\sin \varphi  = 0,\label{ieq}\\
        \varphi_x(0) = h_e - \varkappa_l\,L\gamma,\;\varphi_x(l) = h_e + \varkappa_r\,L\gamma, \label{ibc}
    \end{gather}
Here $L$ is the length of the junction, $h_e$ --- external magnetic field, $\gamma$
--- external current. The parameters $\varkappa_l$ and $\varkappa_r$
($\varkappa_r+\varkappa_r = 1$) characterize the means of injection of the external
current. The existence of the inhomogeneity leads to the local change of the Josephson
current. In this paper the following approximation of the amplitude $j_C(x)$ is used
\begin{eqnarray}
j_C (x) = \left\{ {\begin{array}{*{20}c}
   1 + \kappa, \quad &x \in \Delta, \hfill  \\
   1, &x \notin \Delta. \hfill
 \end{array} } \right.
\end{eqnarray}

Here parameter $\kappa$ describes the portion of Josephson current through the
inhomogeneity. At $\kappa > 0$ we have shunt, at $\kappa \in [-1,0)$  --- microresistive
type of the inhomogeneity. The value $\kappa = 0$ corresponds to the homogeneous
junction.

\begin{figure}
\begin{center}
\epsfxsize=0.7\hsize \epsfig{figure=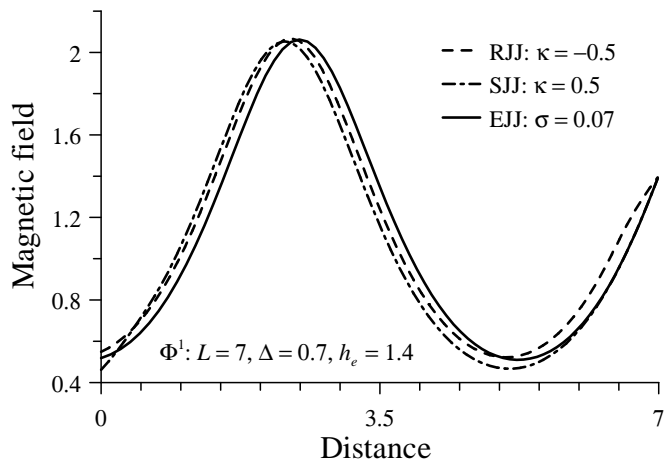,height=6cm} \caption{Distribution of
the internal magnetic field  $\varphi_x(x)$ along the junction for the fluxon $\Phi^1$
at $h_e = 1.4$ and $\lambda_{0} = 10^{-4}$ for three  models EJJ, RJJ and SJJ in the
in-line geometry.} \label{Comp_f1_rse}
\end{center}
\end{figure}

We calculate the bifurcation curves "critical current-external magnetic field" for
static fluxon states and construct the corresponding critical curves for the JJ. The
results of the numerical experiments in the in-line geometry are compared for three
models: exponentially shaped Josephson junction (EJJ) and rectangular Josephson junction
with resistor (RJJ) and shunt (SJJ) inhomogeneities at the ends of junction. We choose
the length of the junction $L = 7$, the width of inhomogeneity $\Delta = 0.7$, and
parameters $\varkappa_l = 1$ and $\varkappa_r = 0$. Fig.\ \ref{Comp_f1_rse} shows the
distribution of the internal magnetic field $\varphi_x(x)$ along the junction for the
fluxon $\Phi^1$ at the value of the external magnetic field $h_e = 1.4$ and the value of
the minimal eigenvalue of the Sturm-Liouville problem $\lambda_{0} = 10^{-4}$, i.e.,
just before the destruction of the fluxon by the external current $\gamma$. We note a
qualitative coincidence of the dependencies  for these three models. The same qualitative
coincidence we have observed for the distribution of the Josephson current. A
quantitative difference takes place near the ends of junction, where the amplitude of
Josephson current in RJJ and SJJ-models is changed. The found values of the critical
currents $\gamma_{cr}(EJJ) \approx 0.126$, $\gamma_{cr}(RJJ) \approx 0.121$,
$\gamma_{cr}(SJJ) \approx 0.134$ at $h_e = 1.4$ are very close to each other.

\begin{figure}
\begin{center}
\epsfxsize=0.7\hsize \epsfig{figure=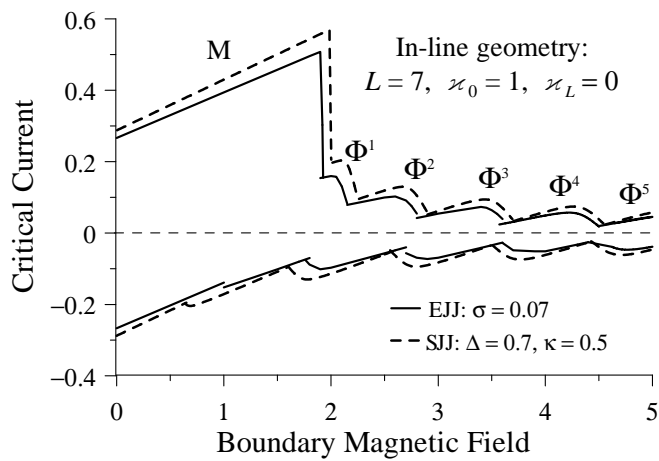,height=6cm} \caption{Critical curves
of the critical current versus magnetic field. The solid curve corresponds EJJ-model
with $\sigma = 0.07$, the dashed curve corresponds to the SJJ-model with a rectangular
shunted inhomogeneity  ($\Delta = 0.7$, $\kappa = 0.5$} \label{Gcrh_sei10}
\end{center}
\end{figure}

In Fig.\ \ref{Gcrh_sei10} we compare the curves ``critical current-magnetic field'' for
EJJ and SJJ models. As we can see they are qualitatively coincide. The similar behaviour
demonstrates the Josephson junction with resistive inhomogeneity at the end of the
junction (RJJ model) \cite{bss_06}.

Both kind of inhomogeneities at the ends of the junction lead to the disappearance of
the mixed fluxon-antifluxon states \cite{tlb_02}. Possibly it explains the experimental
results on the exponentially shaped Josephson junctions concerning the smaller linewidth
of flux flow oscillator, increased output power and better impedance matching to a load.
We consider that the use of the inhomogeneities might have a technological advantage for
construction of the flux flow oscillator.

The authors thank  I.V.Puzinin and N.M.Plakida for useful discussion and cooperation.

\end{document}